\documentclass[11pt]{article}
\usepackage{amsmath}
\usepackage{graphicx}

\begin{document}

\thispagestyle{empty}

\begin{flushright} UCB-PTH-05/19 \end{flushright}
\begin{flushright} LBNL-57959 \end{flushright}

\vskip 0.5cm

\begin{center}{\LARGE { Supersymmetry Then and Now}}

\end{center}

\vskip1cm
\begin{center}
{\bf {Bruno Zumino}}

\vskip0.2cm

Department of Physics, University of California, and\\
Theoretical Physics Group, 50A-5104, Lawrence Berkeley National Laboratory,
Berkeley, CA 94720, USA

\end{center}

\vskip1.0cm

\begin{center}

{\bf ABSTRACT}

\end{center}
\vskip 0.5cm

A brief description of some salient aspects of four-dimensional
supersymmetry: early history, supermanifolds, the MSSM, cold
dark matter, the cosmological constant and the string landscape.

\newpage

\section{A brief history of the beginning of\\
supersymmetry}

Four dimensional supersymmetry (SUSY) has been discovered independently three
times: first in Moscow, by Golfand and Likhtman, then in Kharkov, by Volkov
and Akulov, and Volkov and Soroka, and finally by Julius Wess and me,
who collaborated at CERN in Geneva and in Karlsruhe. It is remarkable
that Volkov and his collaborators didn't know about the work of
Golfand and Likhtman, since all of them were writing papers in Russian
in Soviet journals. Julius and I were totally unaware of the
earlier work. For information on the life and work of Golfand and
Likhtman, I refer to the Yuri Golfand Memorial Volume \cite{Golfand}.
For information on Volkov's life and work, I refer to the Proceedings
of the 1997 Volkov Memorial Seminar in Kharkov \cite{Volkov}.

Supersymmetry is a symmetry which relates the properties of 
integral-spin bosons
to those of half-integral-spin fermions. The generators of the
symmetry form what has come to be called a superalgebra, which is a 
super extension of the Poincar\'e Lie algebra of quantum field theory 
(Lorentz transformations and space-time translations) by 
fermionic generators. In a superalgebra both commutators and 
anticommutators occur. The study of superalgebras is relevant
to the study of dynamical systems with both bosonic and fermionic quantities;
very interesting work on such systems was done in Moscow by the mathematician
F.A. Berezin and his collaborators. It is amusing that one of them,
D.A. Leites, in a book he wrote on the subject, has attributed the
origin of the prefix super to the exaggerated enthusiasm of physicists.
The truth is that, like many other words in physics and mathematics the
technical word ''super'' never had any of the
connotations it has in everyday language.

The work of Golfand and Likhtman and that of Volkov and collaborators 
went to a large extent unnoticed. Instead, the first three preprints
Julius and I wrote aroused immediately the interest of numerious
theoretical physicists, even before publication, and the subject took on a 
life of its own, to which we continued to contribute both together
and separately, with other collaborators. Our early papers also gave
rise to renewed interest by mathematicians in the theory of 
superalgebras. Eventually a complete classification of simple
and semisimple superalgebras was obtained, analogous to Cartan's
classification of Lie algebras, and even the prefix super was adopted
in mathematics. Unfortunately the Poincar\'e  superalgebra is
not semisimple, although it can be obtained by a suitable contraction;
the situation is similar to that for the Poincar\'e Lie algebra.
The general classification of superalgebras does not appear to be
very useful in physics, because, unlike Lie algebras, superalgebras cannot
be used as internal symmetries, or so it seems.\footnote{This is 
true in Minkowski space. Recently the semisimple superalgebra of
$SO(4/2)$ has been used in M-theory (a conjectured eleven
dimensional superstring theory) in the background of $pp$
waves (a solution of Einstein's gravitational equations which 
reduces to Minkowski space in a suitable limit; in the corresponding
limit $SO(4/2)$ reduces to the Poincar\'e superalgebra.)}

The early work on supersymmetric field theories considered only
one fermionic generator which is a Majorana spinor. The
corresponding superalgebra is therefore called $N=1$ SUSY.
An important development was the study of extended
$(N>1)$ SUSY and the construction of quantum field theories
admitting extended SUSY. It turns out that $N=1$ SUSY in four
space-time dimensions is still the best choice for a SUSY
extension of the standard model of elementary particles,
because of the chirality properties of physical fermions.  I
shall describe in a later section a popular version of such an extension, 
the minimal supersymmetric standard model (MSSM).

\section{Remarks on supermanifolds}
The influence of supersymmetry on mathematics can be
seen by the great interest mathematicians have developed in the study
of supermanifolds. From a physicist point of view this began with an
important paper by A. Salam and J. Strathdee who introduced the
concept of ``superspace'', a space with both
commuting and anticommuting coordinates and showed that $N=1$ supersymmetry
can be defined as a set of transformations in superspace.
Wess, Ferrara and I then wrote some papers using the concept of
``superfields'' (fields in superspace). Eventually the
technique of superpropagators was developed and shown to be a 
useful tool for supersymmetric perturbation theory.

With the discovery of supergravity (SUGRA) the supersymmetric extension
of Einstein's gravity) it became natural to study the geometry of
curved supermanifolds. Julius and I realized that the super Riemannian
geometry proposed by R. Arnowitt and P. Nath had to be enlarged by
the introduction of a supervielbein and a constrained, but 
nonvanishing, supertorsion. We also formulated the geometry in terms of
exterior differential superforms, not unlike those introduced independently
by F. Berezin.

\section{The minimal supersymmetric standard model}
Ordinary symmetries of elementary particle physics such as
$SU(3)\otimes SU(2)\otimes U(1)$ arrange particles into
multiplets of different internal quantum numbers but of the
same total spin. Attempts to arrange particles of different
spin in supermultiplets, analogous to Wigner's supermultiplets
in nuclear physics, failed to be consistent with the axioms of
local relativistic quantum field theory. These attempts (such as
``relativistic SU(6)'') involved operators which changed particles
of integral spin into particles of a different integral spin and 
particles of half-integral into particles of a different
half-integral spin. Their failure culminated in the proof of
so called ``no go theorems''. $N=1$ SUSY overcomes these difficulties
by using spin $\frac{1}{2}$ generators which change particle spins by 
$1\over 2$ and their statistics as well. SUSY quantum field
theories are renormalizable theories consistent
with the axioms of relativistic quantum field theory
as is very clear already from the very first papers.
Julius Wess and I wrote. Examples of particles belonging to
a supermultiplet are:
\begin{equation*}
\begin{cases}
\text{GLUON}, &\text{SPIN 1},\quad \text{BOSON} \\[2mm]
\text{GLU{\sl INO}}, &\text{SPIN $\frac{1}{2}$},\quad \text{FERMION}
\end{cases}
\end{equation*}

\begin{equation*}
\begin{cases}
\text{QUARK}, &\text{SPIN $\frac{1}{2}$},\quad \text{FERMION} \\[2mm]
\text{{\sl S}QUARK}, &\text{SPIN 0},\quad \text{BOSON}
\end{cases}
\end{equation*}

It is customary to attach the ending ``ino'' to the 
fermionic superpartner of a boson and the initial ``s'' to 
the bosonic superpartner of a fermion, as indicated above.
Thus the bosonic superpartner of the electron is called selection
and denoted $\tilde e$, the fermionic superpartner of the W meson
is called Wino and denoted $\widetilde W$. It is customary to use a tilde
for the superpartner of a known particle.

The Standard Model (SM) is very successful in describing particle
physics, but some theorists are bothered by the so called ``Hierarchy
Problem'': due to quadratic radiative corrections the mass of the
Higgs scalar would naturally be of order $M_p \sim 10^{18}$GeV, the
Planck mass. A SUSY version of the SM would not have this problem: in
a SUSY Quantum Field Theory (QFT) the quadratic corrections cancel
between boson and fermion loops, as Julius and I noticed in our second
paper; other cancellations also occur.  SUSY {\sl does} solve
the hierarchy problem, but it is historically incorrect to say that it
was ``invented'' to solve the hierarchy problem, as some younger
theorists claim. As explained before, it was invented to have spin
supermultiplets; then Julius Wess and I noticed the cancellation of
divergences, as well as the fact that fewer renormalization constants
are needed in SUSY quantum field theories.

The SM has both boson and fermion fields, but the obvious idea to arrange
them in supermultiplets fails to agree with experiment. It turns out that
one is forced to introduce a new {\sl super}partner field to every
single field present in the SM. {\sl And in addition one must introduce
a second Higgs doublet.}

Let us remember the field content of the SM

\[
\begin{split}
\text{Leptons}: L_i =\binom{v}{e}_{L_i}& =  (1,2, -\frac{1}{2})\\
                     e_{R_i}           & =  (1, 1, -1)\\
\text {Quarks}: Q_i =\binom{u}{d}_{L_i}& =  (3, 2, \frac{1}{6})\\
                     u_{R_i}           & =  (3, 1, \frac{2}{3})\\
                     d_{R_i}           & =  (3, 1, -\frac{1}{3})\\
\text {Higgs }: H = \binom{h^+}{h^0}   &= (1, 2, \frac{1}{2})\\
\end{split}
\]
Here $i=1,2,3$ is the ``family'' index, $L$ and $R$ refer to the
left- and right-handed components of fermions and the numbers in
parenthesis are the $SU(3)\otimes SU(2)\otimes U(1)$ quantum numbers.

Let us now compare the field content of the SM with that of the MSSM.
The rules for building $N=1$ SUSY gauge theories are to assign a
vector superfield (VSF) to each gauge field and a chiral superfield
($\chi$SF) to each matter field. The field content of a VSF is one
gauge boson and a Weyl fermion called gaugino, and of the $\chi$SF
is one Weyl fermion and one complex scalar. The VSF's
transform under the adjoint of the gauge group, while the $\chi$SF's
can be in any representation. Since none of the matter fields of the
SM transform under the adjoint of the gauge group, we cannot identify
them with the gauginos. There are additional constraints dictated by
the chirality and lepton number of the SM fields. The result is that
the minimal choice is to attribute to the $\chi$SF's of the MSSM the
quantum numbers in the table. B and L are baryon and lepton number.
\begin{center}
\begin{tabular}{l c c r r r}
\bigskip
&SU(3) &SU(2) &U(1) &B & L\\
\smallskip
$L_i$ &1&2&$-\frac{1}{2}$ &0 &1\\
\smallskip
$\bar E_i$ &1&1&1&0&-1\\
\smallskip
$Q_i$ &3&2&$\frac{1}{6}$ &$\frac{1}{3}$ &0\\
\smallskip
$\bar U_i$ &$\bar 3$ &1 &$-\frac{2}{3}$ &$-\frac{1}{3}$ & 0\\
\smallskip
$\bar D_i$ &$\bar 3$ &1 &$\frac{1}{3}$&$-\frac{1}{3}$&0\\
\smallskip
$H_1$ &1&2&$-\frac{1}{2}$&0&0\\
\smallskip
$H_2$ &1&2&$\frac{1}{2}$&0&0\\
\end{tabular}
\end{center}
So, one is forced to introduce many new particles, which makes room for
many new interactions not existing in physics, for instance baryon
and lepton number violating interactions. To preserve B and L conservation
(which is automatic in the SM) one introduces R parity conservation. R parity
transforms
\begin{center}
\begin{align}
\text{Particle} \to &\text{Particle}\notag \\
\text{Superpartner} \to &\text{ -- Superpartner} \notag
\end{align}
\end{center}
It is a discrete invariance of the SUSY algebra.

It is well known that, in the SM, the coupling constants of the
strong, electromagnetic and weak interactions run with energy according
to the renormalization group equations to converge (almost) to a 
single value at $\sim 10^{15}$GeV. If one uses the MSSM the 
running of the coupling constants is modified (especially because of the
{\sl two} Higgs {\sl super}multiplets, which together count
as much as six ordinary Higgs fields) and they converge much better,
now at $\sim 10^{16}$GeV; is this a hint at unification with
supergravity?\\

\noindent {\bf SUSY breaking (the hidden sector)}\\
Exact SUSY implies that a particle and its superpartner have the same
mass, which is clearly not true in the real world. So SUSY must be
broken but not too violently in order not to lose the desirable features
of SUSY quantum field theories, such as the cancellation of quadratic
divergences, the unification of couplings etc. Like other symmetries
SUSY can be broken ``spontaneously'', which would satisfy that
requirement; however spontaneous breaking still preserves relations among
the masses (mass sum rules) which are not satisfied in reality. So one
must find some other way to break SUSY ``softly''.

A popular approach is to postulate the existence of a ``hidden sector'' where 
SUSY is broken spontaneously at much higher energy scales than the weak
scale. The sector is hidden in the sense that its fields do not
interact with the SM particles (``visible sector'' except through
``minimal'' supergravity which will mediate the SUSY breaking to the
visible sector. The idea of a hidden sector may seem far fetched, but
it emerges naturally in some versions of superstring theory, e.g. heterotic
string theory.

\section{The cold dark matter problem}

Many independent lines of cosmological evidence have led to the conclusion
that the vast majority of matter in the universe is ``dark'' (it has evaded
observation based on direct interaction with electromagnetic radiation).
Nonbaryonic dark matter out-masses the ordinary matter by a factor
of approximately 8. The dominant class of dark matter candidates are
``Weakly Interacting Massive Particles'' (WIMPS).  There have been a 
number of suggestions for dark matter particles but it seems that the best
candidate is provided by TeV-scale SUSY as the ``neutralino''.

A particle dark matter candidate must satisfy the following criteria:
\begin{itemize}
\item It must be ``stable'' (long lifetime compared to the age of the universe)
to contribute to structure formation.

\item
There must be an effective production mechanism to create the right
amount in the early universe.

\item
It must be ``nonrelativistic'' during structure formation (``cold'' dark 
matter).

\item
It must be ``weakly interacting'' to have escaped detection, 
electrically neutral and colorless.
\end{itemize}

These constraints are satisfied by the neutral Higgsinos
$(\tilde H_u,\tilde H_d)$, the neutral Wino $(\tilde W^0)$ and
the Bino $(\tilde B^0)$, four Majorana fermions with the
same quantum numbers, which can mix giving four mass eigenstates, the
neutralinos $\chi^0_1,\chi^0_2,\chi^0_3,\chi^0_4$. The lightest one is
a good candidate.

Other possibilities are: the lightest mixing of sneutrinos (apparently
excluded by accelerator searches) or the gravitino (superpartner of the
graviton, very hard to detect).

Depending on the model of SUSY SM, we are talking about the
{\sl lightest superpartner (LSP)}.

\section{The cosmological Constant, or Dark Energy, and the Vacuum
Energy Problem.}

It appears to be generally accepted, by astrophysicists and cosmologists,
that the cosmological constant $\Lambda$ (long believed to be zero)
is actually positive but very small. As a consequence, the
expansion of the universe accelerates. If we interpret $\Lambda$
as the energy of the vacuum, dimensional arguments, as well as
quantum field theory (QFT) calculations would give it a value of
\begin{equation}
\Lambda_p = \frac{\text{Planck Mass}}{\text{(Planck Length)}^3}\approx
10^{94} \frac{\text{grams}}{\text{cm}^3}\notag
\end{equation}
The actual value is $\Lambda\sim 10^{-120}\Lambda_p$.

In a SUSY QFT (without SUGRA), the vacuum energy vanishes to all
orders in perturbation theory. In a generic QFT it diverges
quadratically while if SUSY is broken only softly it diverges 
logarithmically. Still, for any reasonable cut-off, $\Lambda$
comes out much larger than the above measured value.
So, SUSY does not seem to explain the smallness of $\Lambda$.
Recently, within the framework of supersymmetric string theory an
approach to this problem has emerged, which has been named
the ``String Theory Landscape'' (Bousso, Polchinski,
Susskind, Douglas and others) and which makes use of the so-called 
``Anthropic Principle'', in a form discussed some time ago by
S. Weinberg.

Let us accept that the basic equations of superstring theory are
{\sl given}, in ten dimensions. These equations have many solutions and
one is interested in those where six dimensions are compactified (most
compactifications studied are in Calabi-Yau manifolds). the
resulting theory in four dimensions depends on the topology of the
manifold and on the values of various string theory fluxes of
fields wrapped around handles of the manifold. Some string theorists
count up to 500 handles and different numbers of flux lines (0 to 9). So,
one could have $100^{500}$ parameters upon which physics, and the vacuum 
energy, in four dimension can depend (see figure). The vacuum energy
for each valley corresponds to the local minimum, each valley
corresponds to a stable (or metastable) set of physical laws.
The figure assumes only one parameter: the size of the
compact manifold. The true string theory landscape reflects all
parameters and forms a topography with a vast number of dimensions.
The entire visible universe exists within a region of space
associated with a valley that happens to produce laws of physics
suitable for the evolution of life. Weinberg considered a restricted
form of the anthropic principle, in which one assumes that all constants
of nature (e.g.\ the time structure constant, mass ratios of elementary
particles etc) have the observed values and only the constant $\Lambda$
is arbitrary. He showed that this requires a $\Lambda$ very close to
the observed value, otherwise galaxies would not have formed.

\begin{figure}[tb]
\centering
\includegraphics[width=.95\textwidth]{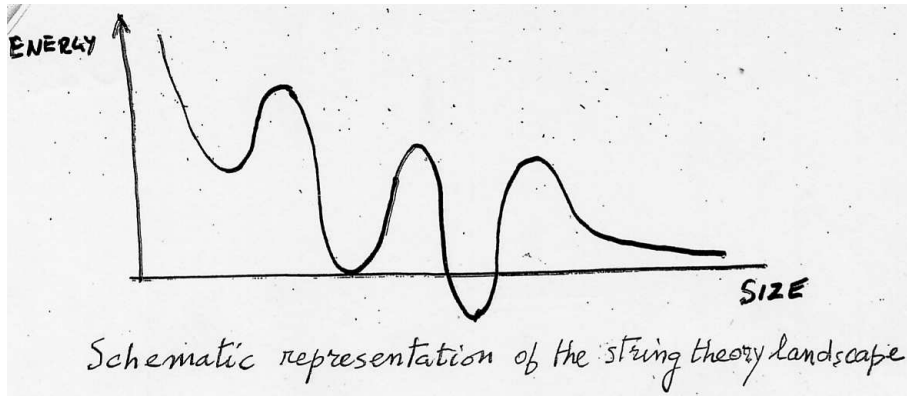}
\caption{Schematic representation of the string theory landscape}
\end{figure}

\section{Concluding remarks}
In the present paper, which is based on a colloquium lecture,
I necessarily had to limit myself to a very sketchy description of a few
topics. For the reader who is interested in a deeper understanding, I 
can recommend some papers, books and review articles.

For Supersymmetry, Supergravity and Superstring theory I refer to
\cite{WB, Weinberg, FF, PvanN, Polchinski}. A very clear description of
the MSSM can be found in \cite{Csaki}.  For the cold dark matter problem, 
there is a very comprehensive recent review \cite{BHS}. For the
cosmological constant problem and the anthropic principle see
\cite{SWeinberg}; comprehensive reviews are \cite{PR, Pad}.

Before concluding, it should be mentioned that some theorists have
argued that one should not worry too much about the hierarchy problem,
which they consider a merely ``philosophical'' or ``aesthetic''
matter. This gives them the freedom to fine tune parameters
arbitrarily and thus to invent new models (see H. Murayama's ``New
SM'').  If the SM is only an effective field theory, this point of
view, which is at variance with arguments given above, is perhaps not
totally unreasonable; however, it does not seem to have gained much
acceptance.

\section*{Acknowledgement}
I am very grateful to Jeff Filippini, who made available to me his term paper
on ``The Direct Detection of Supersymmetric Dark Matter''. This work
was supported in part by the Director, Office of Science, Office
of High Energy and Nuclear Physics, of the U.S. Department of Energy
under contract No. DE-AC03-76SF00098, and in part by the NSF under Grant
No. PHY-0098840.


\begin{thebibliography}{0}
\bibitem{Golfand} Yuri Golfand Memorial Volume, M. Schifman, ed. (World
   Scientific, Singapore, 2000).
\bibitem{Volkov} D. Volkov Memorial Seminar, J. Wess and V.P. Akulov, eds.
   (Springer, Berlin, 1998).
\bibitem{WB} J. Wess and J. Bagger, Supersymmetry and Supergravity,
   2nd ed. (Princeton University Press, Princeton, NJ, 1992).
\bibitem{Weinberg} S. Weinberg, The Quantum Theory of Fields, vol. III,
   Supersymmetry (Cambridge University Press, Cambridge, 2000).
\bibitem{FF} P. Fayet and S. Ferrara, Physics Reports {\bf 32 C}, no. 5,
   249-334 (1977).
\bibitem{PvanN} P. van Nieuwenhuizen, Physics Reports {\bf 68}, no. 4,
   189-398 (1981).
\bibitem{Polchinski} J. Polchinski, String Theory (Cambridge University Press,
   Cambridge, 1988) (two volumes).
\bibitem{Csaki} C. Csaki, Mod. Phys. Lett. {\bf A11}, 599-613 (1996).
\bibitem{BHS} G. Bertone, D. Hooper, J. Silk, Physics Reports {\bf 405},
   279-390 (2005).
\bibitem{SWeinberg} S. Weinberg, Phys. Rev. Lett. {\bf 59}, 2607 (1987);\\
   Rev. Mod. Phys. {\bf 61}, 1 (1989); in Relativistic Astrophysics,
   J.C. Wheler and H. Martel eds., AIP Conf. Proc. No 586 (AIP, Melville, NY),
   p. 893 (2001).
\bibitem{PR} P.J.E. Peebles, B. Ratra, Rev. Mod. Phys. 
   {\bf 75}, 559-606 (2003).
\bibitem{Pad} T. Padmanabhan, Physics Reports {\bf 380}, 235-320 (2003).
\end{thebibliography}
\end{document}